\documentclass[12pt]{article}
\usepackage{a4wide,epsfig,psfrag,amsmath,amssymb,cite,scalefnt}
\usepackage{color}
\usepackage{amsmath,comment,braket}

\graphicspath{ {figs/} }

\newcommand{\be}{\begin{equation}}
\newcommand{\ee}{\end{equation}}
\newcommand{\bea}{\begin{eqnarray}}
\newcommand{\eea}{\end{eqnarray}}

\allowdisplaybreaks

\newcommand{\gsim}{\;\rlap{\lower 3.5 pt \hbox{$\mathchar \sim$}} \raise 1pt
 \hbox {$>$}\;}
\newcommand{\lsim}{\;\rlap{\lower 3.5 pt \hbox{$\mathchar \sim$}} \raise 1pt
 \hbox {$<$}\;}

\begin{document}

\title{\vskip-3cm{\baselineskip14pt
    \begin{flushleft}
      \normalsize 
      DESY 19-092\\
      HU-EP-19/13\\
      P3H-19-012\\
      TTP19-014\\
  \end{flushleft}}
  \vskip1.5cm
  Top quark mass dependence of the Higgs-gluon form factor at three loops
}

\author{
  Joshua Davies$^{(a)}$, 
  Ramona Gr\"ober$^{(b)}$, 
  Andreas Maier$^{(c)}$, 
  \\
  Thomas Rauh$^{(d)}$, 
  Matthias Steinhauser$^{(a)}$,
  \\[1mm]
  {\small\it (a) Institut f{\"u}r Theoretische Teilchenphysik,}\\
  {\small\it Karlsruhe Institute of Technology (KIT)}\\
  {\small\it Wolfgang-Gaede Stra\ss{}e 1, 76128 Karlsruhe, Germany}
  \\[1mm]
  {\small\it (b) Humboldt-Universit\"at zu Berlin, Institut f\"ur Physik,}\\
  {\small\it Newtonstr. 15, 12489 Berlin, Germany}
  \\[1mm]
  {\small\it (c) Deutsches Elektronen-Synchrotron, DESY,}\\
  {\small\it Platanenallee 6, 15738 Zeuthen, Germany}
  \\[1mm]
  {\small\it (d) Albert Einstein Center for Fundamental Physics,}\\
  {\small\it Institute for Theoretical Physics, University of Bern,}\\
  {\small\it Sidlerstrasse 5, CH-3012 Bern, Switzerland}
  }

\date{}

\maketitle

\thispagestyle{empty}

\begin{abstract}

  We compute three-loop corrections to the Higgs-gluon form factor,
  incorporating the top quark mass dependence. Our method is based on the
  combination of expansions around the top threshold and for large top
  quark mass, using conformal mapping and Pad\'e approximation to describe the
  form factor over the full kinematic range.

\end{abstract}

\thispagestyle{empty}

\sloppy


\newpage


\section{Introduction}

The precise measurement of the properties of the Higgs boson, in particular the
coupling strength to other particles and to itself, will be among the main
focuses in particle physics in the coming years. The success of this
enterprise crucially depends on the accuracy of the predictions provided by
the theory community.

A quantity which is available to high perturbative order is the total cross
section for the production of a Higgs boson at the Large Hadron Collider
(LHC).  For a comprehensive collection of relevant works we refer to
Ref.~\cite{deFlorian:2016spz}, but we remark here that QCD corrections
including the exact dependence on the top quark mass, $m_t$, have been
available at next-to-leading order (NLO) for about 25
years~\cite{Spira:1995rr}. At higher orders only approximate results are
available; at NNLO the infinite top quark mass results from
Refs.~\cite{Harlander:2002wh,Anastasiou:2002yz,Ravindran:2003um} have been
complemented by power-suppressed terms in the inverse top quark mass
in~\cite{Harlander:2009mq,Pak:2009dg,Harlander:2009my,Pak:2011hs}.  The
N$^3$LO result has been obtained in the $m_t \to \infty$ limit
in~\cite{Anastasiou:2016cez,Mistlberger:2018etf}.

In Ref.~\cite{deFlorian:2016spz} several sources of uncertainties have been
identified for the prediction of the total cross section. Among them is that
of the exact top quark mass dependence of the NNLO corrections which has been
estimated to be 1\%. In this paper we provide results for the Higgs-gluon form
factor at three-loop order which constitutes the virtual corrections to the
production cross section. Thus the findings of this paper help to eliminate
the aforementioned uncertainty to a large extent.
The Higgs-gluon form factor is also an important ingredient 
for processes where the relevant energy in the fermion loops reaches values
close to or above the top quark threshold and the infinite top quark mass
limit cannot be applied anymore. This concerns, e.g.,
Higgs boson pair production via $gg\to H^\star \to HH$
or the measurement of the Higgs boson width from off-shell production of $Z$
boson pairs in gluon fusion via $gg\to H^\star \to ZZ$~\cite{Caola:2013yja}.
The exact dependence
on the fermion mass in the loop is also important for numerous theories beyond
the Standard Model, which often contain additional heavier Higgs bosons.

At two-loop order exact results for the form factor are known from
Refs.~\cite{Spira:1995rr,Harlander:2005rq,Anastasiou:2006hc,Aglietti:2006tp}.
However, at three loops only expansions for large top quark
mass~\cite{Harlander:2009bw,Pak:2009bx} and non-analytic terms in the
expansion around the top threshold up to ${\cal O}(1-z)$~\cite{Grober:2017uho}
are known, where
\begin{equation}
	z=\frac{\hat{s}}{4m_t^2}
	\label{eq::zdefn}
\end{equation}
with $\sqrt{\hat{s}}$ being the partonic center-of-mass energy.
For later convenience  we also introduce $\bar{z}=1-z$.
In the next section we describe our method which we
use to combine these expansions in order to obtain results for the form factor
valid for all space- and time-like momentum transfers.  In
Section~\ref{sec::res} we discuss our results and Section~\ref{sec::con}
contains a brief summary.


\section{Method}

The method we use for the construction of the top quark mass dependence of the
Higgs-gluon form factor is based on the efficient combination of information
from the large top quark mass expansion (LME) ($z\to 0$) and knowledge from
the threshold where $\hat{s}\approx 4m_t^2$ ($z\to 1$), using conformal mapping
and Pad\'e approximation.  The procedure was developed in
Ref.~\cite{Baikov:1995ui} (see
also~\cite{Broadhurst:1993mw,Fleischer:1994ef}) in order to compute a
certain class of four-loop contributions to the muon anomalous magnetic
moment. In Refs.~\cite{Chetyrkin:1995ii,Chetyrkin:1998ix} the method was
extended to QCD corrections with the aim to compute NNLO correction to the
total cross section $\sigma(e^+e^-\to\mbox{hadrons})$. A further refinement of
the method has been developed in Refs.~\cite{Hoang:2008qy,Kiyo:2009gb} where
order $\alpha_s^3$ corrections to $\sigma(e^+e^-\to\mbox{hadrons})$ have been
computed. In these references additional parameters were introduced which
allow one to generate a larger number of Pad\'e approximations and thus
obtain more reliable uncertainty estimates.  The systematic improvement
  of the Pad\'e approximations when increasing the number of input terms has
  been studied in Ref.~\cite{Maier:2017ypu}.  In Ref.~\cite{Grober:2017uho}
the method has been used to obtain two-loop corrections for the three form
factors relevant for Higgs boson pair production.

\begin{figure}[t]
  \centering
  \includegraphics[width=\textwidth]{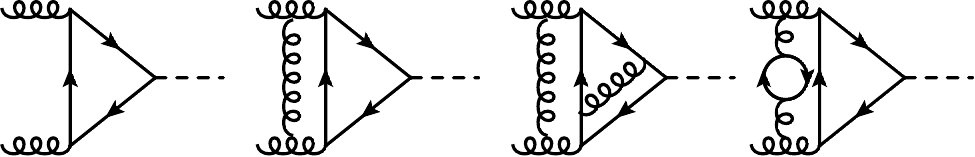}
  \caption{\label{fig::diags}One-, two- and three-loop Feynman diagrams
    contributing to $F_\triangle$. Solid, curly and dashed lines represent
    quarks, gluons and Higgs bosons, respectively.}
\end{figure}

In the following we briefly describe the application of this method to the form factor
entering the interaction of a Higgs boson and two gluons. We parameterize the 
corresponding amplitude as
\begin{eqnarray}
  {\cal A}^{\mu\nu}_{ab}(gg \to H)
  &=& \delta_{ab} \frac{y_t}{2\sqrt{2}m_t} \frac{\alpha_s}{\pi}
       T_F 
      \left(q_1\cdot q_2 g^{\mu\nu} - q_1^\nu q_2^\mu\right)
      F_\triangle(z)
      \,,
      \label{eq::amp}
\end{eqnarray}
where $q_1$ and $q_2$ are the external momenta of the gluons with polarization
vectors $\varepsilon^\mu(q_1)$ and $\varepsilon^\nu(q_2)$,
respectively. $y_t=\sqrt{2}m_t/v$ is the top quark Yukawa coupling, $v$ is the
vacuum expectation value, $a$ and $b$ are adjoint colour indices,
$T_F=1/2$ and $\hat{s}=(q_1+q_2)^2=2q_1\cdot q_2$. It is convenient to define
the perturbative expansion of $F_\triangle$ as
\begin{eqnarray}
  F_\triangle &=& F_\triangle^{(0)} + \frac{\alpha_s}{\pi} F_\triangle^{(1)}
               + \left(\frac{\alpha_s}{\pi}\right)^2 F_\triangle^{(2)} + \cdots
               \,,
\end{eqnarray}
where $\alpha_s \equiv \alpha_s^{(5)}(\mu)$ is the strong coupling
constant with five active flavours evaluated at the renomalization scale $\mu$.
Sample Feynman diagrams contributing to ${\cal A}^{\mu\nu}_{ab}(gg \to H)$
up to three loops can be found in Figure~\ref{fig::diags}.

The one-loop result, $F^{(0)}_\triangle$, is finite.  At two-loop order
we renormalize the gluon wave function and the top quark mass in the
on-shell scheme and the strong coupling constant in the $\overline{\rm MS}$ scheme.
Note that the ultra-violet renormalized form factor still contains infra-red
divergences which cancel against contributions from real radiation, in order
to form finite physical quantities. The structure of the infra-red divergences
is universal and has been studied in detail in the
literature~\cite{Catani:1998bh}. In our case finite form factors are obtained
via
\begin{eqnarray}
  F^{(1),\rm fin}_\triangle &=& F_\triangle^{(1)} - \frac{1}{2} I^{(1)}_g F_\triangle^{(0)}\,,\nonumber\\
  F^{(2),\rm fin}_\triangle &=& F_\triangle^{(2)} - \frac{1}{2} I^{(1)}_g F_\triangle^{(1)}
  - \frac{1}{4} I^{(2)}_g F_\triangle^{(0)}\,,
  \label{eq::FF_IR}
\end{eqnarray}
where $I^{(1)}_g$ and $I^{(2)}_g$ can be found in
Refs.~\cite{Catani:1998bh,deFlorian:2012za}. In order to fix the notation
we provide an explicit expression only for $I^{(1)}_g$ which is given by
\begin{eqnarray}
	I^{(1)}_g &=&
		{} - \left(\frac{\mu^2}{- \hat{s} -i\delta}\right)^\epsilon
		\frac{e^{\epsilon\gamma_E}}{\Gamma(1-\epsilon)}
		\frac{1}{\epsilon^2}
		\Big[
			C_A + 2\epsilon\beta_0
		\Big]\,,
\end{eqnarray}
with $\beta_0 = (11C_A - 4 T_Fn_l)/12$ where $C_A=3$, $T_F=1/2$ and $n_l$
is the number of massless quarks. We
work in $d=4-2\epsilon$ dimensions and assume that $\delta$ is an
infinitesimal small parameter.
We apply
the method described below to $F^{(1),\rm fin}_\triangle$ and
$F^{(2),\rm fin}_\triangle$.

In the following we briefly discuss the input for the limits $z\to0$ and $z\to1$ used for
the construction of the Pad\'e approximants. For the renormalization scale we
choose $\mu^2=-\hat{s}$ since the $\mu$ dependence can easily be reconstructed from
the one- and two-loop expressions, which are known exactly, see Ref.~\cite{DavSte}.
Furthermore, we set all colour factors to their numerical values and only keep
$n_l$ as a parameter. The large-$m_t$ expansion
of the three-loop form factor up to order $z^4$ has been computed in
Ref.~\cite{Harlander:2009bw,Pak:2009bx} and the $z^5$ and $z^6$ terms are
available from Ref.~\cite{DavSte}. The analytic expressions read
\begin{eqnarray}
   F_{\triangle}^{(0)} &=& 
  \frac{4}{3}
 + \frac{14}{45} z
 + \frac{8}{63} z^2
 + \frac{104}{1575} z^3
 + \frac{2048}{51975} z^4
 + \frac{4864}{189189} z^5
 + \frac{512}{28665} z^6
  + {\cal O}(z^7)\,, \nonumber\\ \vphantom{\Bigg[}
  F_{\triangle}^{(1),\rm fin} &=&                                   
  \frac{11}{3}
 + \frac{1237}{810} z
 + \frac{35726}{42525} z^2
 + \frac{157483}{297675} z^3
 + \frac{2546776}{7016625} z^4
 + \frac{194849538824}{737482370625} z^5
                                  \nonumber\\&&\mbox{}
 + \frac{385088204192}{1917454163625} z^6
  + {\cal O}(z^7)\,, \nonumber\\ \vphantom{\Bigg[}
  F_{\triangle}^{(2),\rm fin} &=&
      -\frac{253 \zeta (3)}{24}+\frac{3941}{108}+\frac{19 \pi ^2}{12}+\frac{\pi ^4}{96}    
+      \frac{19}{12}    L_s
\nonumber\\ &&\mbox{}
  + n_l\Bigg(
      -\frac{17 \zeta (3)}{36}-\frac{3239}{648}-\frac{47 \pi ^2}{432}    
+      \frac{4}{9}    L_s
  \Bigg)
 + n_l^2      \frac{\pi ^2}{648}\nonumber\\ &&\mbox{}
\nonumber\\ &&\mbox{}+\Bigg[
      \frac{9290881 \zeta (3)}{103680}-\frac{44326367}{466560}+\frac{623 \pi ^2}{1080}+\frac{7 \pi ^4}{2880}+\frac{28}{405} \pi ^2 \log (2)    
+      \frac{8261}{3240}    L_s
\nonumber\\ &&\mbox{}  + n_l\Bigg(
      -\frac{119 \zeta (3)}{1080}-\frac{107087}{291600}-\frac{259 \pi ^2}{4320}    
      -\frac{169}{1080}    L_s
  \Bigg)
  + n_l^2      \frac{7 \pi ^2}{19440}\Bigg]
 z
\nonumber\\ &&\mbox{}+\Bigg[
      \frac{7037623781 \zeta (3)}{69672960}-\frac{82500975779}{731566080}+\frac{121 \pi ^2}{378}+\frac{\pi ^4}{1008}+\frac{32}{567} \pi ^2 \log (2)    
\nonumber\\ &&\mbox{}
+      \frac{253549}{170100}    L_s
  + n_l\Bigg(
      -\frac{17 \zeta (3)}{378}-\frac{6385481}{53581500}-\frac{25 \pi ^2}{648}    
     -\frac{4133}{36450}    L_s
  \Bigg)
\nonumber\\ &&\mbox{}
 + n_l^2      \frac{\pi ^2}{6804}\Bigg]
 z^2
+\Bigg[
      \frac{650760513719 \zeta
               (3)}{412876800}-\frac{1740869750908152049}{921773260800000}+\frac{221
               \pi ^2}{1050}
\nonumber\\ &&\mbox{}
+\frac{13 \pi ^4}{25200}
+\frac{208 \pi ^2 \log (2)}{4725}    
+      \frac{804644}{826875}    L_s
  + n_l\Bigg(
      -\frac{221 \zeta (3)}{9450}-\frac{6383750249}{112521150000}
\nonumber\\ &&\mbox{} 
-\frac{3107 \pi ^2}{113400}    
      -\frac{1147037}{14883750}    L_s
  \Bigg)
 + n_l^2      \frac{13 \pi ^2}{170100}\Bigg]
 z^3
\nonumber\\ &&\mbox{}+\Bigg[
      \frac{193543938976537 \zeta
               (3)}{37158912000}-\frac{6978205934887756008911}{1115345645568000000}+\frac{4736
               \pi ^2}{31185}+\frac{16 \pi ^4}{51975}
\nonumber\\ &&\mbox{}
+\frac{16384 \pi ^2 \log (2)}{467775}    
+      \frac{33498106}{49116375}    L_s
+ n_l\Bigg(
      -\frac{2176 \zeta (3)}{155925}-\frac{2197298833}{72937816875}
\nonumber\\ &&\mbox{}
-\frac{3232 \pi ^2}{155925}    
      -\frac{20932}{382725}    L_s
  \Bigg)
+ n_l^2      \frac{64 \pi ^2}{1403325}\Bigg]
 z^4
\nonumber\\ &&\mbox{}+\Bigg[
      \frac{2460310706266276921 \zeta
               (3)}{81155063808000}-\frac{159929147625953730170902566067}{4389031448658778521600000}
\nonumber\\ &&\mbox{}
+\frac{9424 \pi ^2}{81081}+\frac{38 \pi ^4}{189189}+\frac{48640 \pi ^2 \log (2)}{1702701}    
+      \frac{945911804923}{1877227852500}    L_s
+ n_l\Bigg(
      -\frac{5168 \zeta (3)}{567567}
\nonumber\\ &&\mbox{} 
-\frac{22552503119716043}{1395235522161731250}-\frac{27892 \pi ^2}{1702701}    
      -\frac{48324340168}{1191317675625}    L_s
  \Bigg)
+ n_l^2      \frac{152 \pi ^2}{5108103}\Bigg]
 z^5
\nonumber\\ &&\mbox{}+\Bigg[
      \frac{15128773883548934558969 \zeta
               (3)}{114266329841664000}
+\frac{2656 \pi ^2}{28665}+\frac{4 \pi ^4}{28665}+\frac{2048 \pi ^2 \log (2)}{85995}    
\nonumber\\ &&\mbox{} \vphantom{\Bigg[}
-\frac{13560383230749413568271118392175429}{85205730523295753699328000000}
+      \frac{339242844181}{871570074375}    L_s
\nonumber\\ &&\mbox{} \vphantom{\Bigg[}
 + n_l\Bigg(
      -\frac{544 \zeta (3)}{85995}
-\frac{2085146760850288}{259115168401464375}-\frac{3448 \pi ^2}{257985}    
      -\frac{35895528824}{1150472498175}    L_s
  \Bigg)
\nonumber\\ &&\mbox{} 
+ n_l^2      \frac{16 \pi ^2}{773955}\Bigg]
 z^6
  + {\cal O}(z^7)\,,
      \label{eq::Flme}
\end{eqnarray}
where $L_s = \log(-4z-i0)$ and $\zeta(n)$ is the Riemann zeta
function.

The expansion of the three-loop form factor around threshold was
considered in Ref.~\cite{Grober:2017uho} in the effective theory of
Non-Relativistic QCD (NRQCD). We briefly outline the approach and
refer to~\cite{Grober:2017uho} for details.

Within NRQCD, the leading contributions to the triangle form factor
near threshold can be written schematically as
\begin{equation}
  \label{eq:thr_exp}
  F_{\triangle}(z) \mathop{\asymp}\limits^{z\to1} C_{gg \to t\bar{t}}
  C_{t\bar{t} \to H} G_P(z)\,,
\end{equation}
where the symbol ``$\asymp$'' indicates that terms analytic in $(1-z)$
have been dropped on the right-hand side. The relativistic
short-distance corrections to top pair production and annihilation are
absorbed into the matching coefficients $C_{gg \to t\bar{t}},
C_{t\bar{t} \to H}$ and the propagation of the intermediate
non-relativistic $t\bar{t}$ pair is described by the Coulomb resummed
$P$-wave Green function $G_P(z)$~\cite{Beneke:2013kia}. Expanding
Eq.~\eqref{eq:thr_exp} in $\alpha_s$ yields the perturbative
coefficients of the form factor. An explicit result for the three-loop
form factor is given in Eq.~(50) of~\cite{Grober:2017uho}. For
convenience we reproduce the analytic expression together with the
one- and two-loop results which are given by
\begin{align}
 F_\triangle^{(0)} \mathop{\asymp}\limits^{z\to1} & \,
     2 \pi (1-z)^{3/2}
   + \frac{13 \pi}{3} (1-z)^{5/2}
   + \mathcal{O}\left((1-z)^{7/2}\right)\,, \nonumber\\
 F_\triangle^{(1),\text{fin}} \mathop{\asymp}\limits^{z\to1} & \,
     \frac{4 \pi ^2}{3}(1-z)\log(1-z)
   - \frac{\pi}{36} \left(124+15 \pi ^2\right) (1-z)^{3/2}
   + \frac{8 \pi ^2}{9} (1-z)^2 \log(1-z) \nonumber\\
 & + \frac{\pi}{216} \Big[2252-117 \pi ^2-2112 \log (2)-672
\log(1-z)\Big] (1-z)^{5/2} \nonumber\\
 & - \frac{28\pi ^2}{45} (1-z)^3 \log(1-z)
   + \mathcal{O}\left((1-z)^{7/2}\right)\,, \nonumber\\
 F_\triangle^{(2),\text{fin}} &\mathop{\asymp}\limits^{z\to1} 
  - \frac{8\pi^3}{27} \left(3+\pi ^2\right) \sqrt{1-z} \nonumber\\
  & + \frac{\pi^2}{54}\Big[\left(458-15 \pi^2-44 n_l
    {+ (198-12n_l)(L_s-2\ln2) } \right)\log(1-z)\nonumber\\
    & - (99-6 n_l) \log^2(1-z)\Big](1-z)
  + \mathcal{O}\left((1-z)^{3/2}\right)\,.
      \label{eq::Fthr}
\end{align}

The information provided in Eqs.~(\ref{eq::Flme}) and~(\ref{eq::Fthr})
is used to construct approximations of the form factor.
We first subtract the logarithmic contributions for $z\sim 1$ and define
\begin{equation}
  \tilde{F}_\triangle = F_\triangle - F_\triangle^{\rm sub}
  \,,
  \label{eq::Fthrsub}
\end{equation}
where $F_\triangle^{\rm sub}$ is constructed to both be analytic for
$|z| < 1$ and to reproduce the threshold logarithms in
Eq.~(\ref{eq::Fthr}), so that the threshold expansion of
$\tilde{F}_\triangle$ is free of logarithms up to $(1-z)^{3/2}$.
Such a subtraction function $F_\triangle^{\rm sub}$ can be
obtained using the vacuum polarization as a building block,
see~\cite{Grober:2017uho} for details of the construction.
For explicit examples for $F_\triangle^{\rm sub}$ we refer to the sample Pad\'e approximants in the ancillary file~\cite{progdata}.
Note that also in the limit $z\to0$ $F_\triangle$ develops logarithmic
divergences which manifest in the linear $L_s$ term in Eq.~(\ref{eq::Flme}).
Whereas in Ref.~\cite{Chetyrkin:1998ix} these contributions are also
subtracted, here we instead construct separate Pad\'e approximants for the
$L_s$-independent term and for the coefficient of $L_s$,
as discussed in~\cite{Grober:2017uho}.

Next we apply a conformal mapping
\begin{eqnarray}
  z &=& \frac{4\omega}{(1+\omega)^2}
        \,,
\end{eqnarray}
to transform the $z$ plane into the interior of the unit circle in the
$\omega$ plane; the time-like momentum regions $z\in[0,1]$ and
$z\in[1,\infty]$ with $\mbox{Im}(z)>0$ are mapped to $\omega\in[0,1]$
and the upper semi-circle, respectively.

At this point we construct Pad\'e approximants in the variable $\omega$.
They have the form
\begin{equation}
  [n/m](\omega) = \frac{\sum\limits_{i=0}^n a_i \omega^i}{1 + \sum\limits_{j=1}^m b_j \omega^j}
  \label{eq:Pade_ansatz}
  \,,
\end{equation}
where $n+m$ is fixed by the number of input terms from the large top mass and
threshold expansions. In our case we have seven terms for $z\to 0$ and
one for $z\to1$ which is sufficient to determine eight coefficients in
Eq.~(\ref{eq:Pade_ansatz}), i.e. Pad\'e approximants for $n+m=7$.
More precisely, we construct Pad\'e approximants for the rescaled form factor 
\begin{equation}
  [n/m](\omega) \simeq \left[1+a_R\,z(\omega)\right] \tilde{F}_\triangle(z(\omega)),
  \label{eq:Pade_rescaling}
\end{equation}
where $a_R$ is a real parameter. This removes the spurious condition
$F_\triangle(z\to\infty)=0$ introduced by the definition of the form factor
through $\mathcal{A}_{gg\to H}\propto zF_\triangle(z)$ and provides a means to
test the stability of the solutions through variation of $a_R$.  As
discussed in~\cite{Grober:2017uho} we only use the diagonal and
next-to-diagonal Pad\'e approximants which are $[5/2],[4/3],[3/4]$ and $[2/5]$
in the case that
seven large top quark mass expansion terms and one term from the threshold
expansion are taken into account. In Section~\ref{sec::res} we also show
results which only incorporate LME terms up to $z^4$, for which we construct
the Pad\'e approximants $[4/1],[3/2],[2/3]$ and $[1/4]$.

By construction the Pad\'e approximants develop poles in the complex
$\omega$ plane. In the following we discuss our criteria which exclude
approximants with poles too close to the physical region.
For this discussion we have to distinguish space-like and time-like momentum regions.
For $z>0$ we exclude all approximants which contain poles $\omega_0$ in the region 
\begin{equation} 
  \text{Re}(z(\omega_0)) \geq -2 
  \quad\&\quad
  |\omega_0|\leq1.2\,,
  \label{eq:pole_crit}
\end{equation}
as they can cause unphysical behaviour in the approximation.  We find that
poles in the entire complex plane in $z$, i.e. in the unit disc
$|\omega|\leq1$, cannot be excluded as this leads to the exclusion of all
Pad\'e approximants. In one case we moderately relaxed the exclusion
  region in order to find a sufficient number of Pad\'e approximants. This
  concerns the non-$L_s$ term $F_\Delta^{(2),\text{fin}}(z)$ which we split
  into the $n_h$ contribution with an additional heavy-quark loop and the
  remainder. For the $n_h$ term we relax the exclusion region to
\begin{equation}
 \text{Re}(z(\omega_0)) \geq -1 \qquad \& \qquad |\omega_0| \leq 1.2
 \label{eq:pole_crit_2}
\end{equation}
to find a sufficient number of Pad\'e approximants. The residues of the
poles in the region $-2 \leq \text{Re}(z(\omega_0)) \leq -1$ are typically three
orders of magnitude smaller than those in the remaining part of the complex
plane and therefore do not cause any visible unphysical resonances.

For each choice of $[n/m]$ we aim to construct 20 Pad\'e approximants by
choosing $a_R$ in Eq.~(\ref{eq:Pade_rescaling}) randomly in the range
$[0.1,10]$, leading to a maximum of 80 approximants. The mean and standard
deviation of this set are used as the central value and uncertainty
estimate, respectively. For some choices of $\{n,m\}$ Pad\'e approximants
satisfying criteria ~(\ref{eq:pole_crit}) and~(\ref{eq:pole_crit_2}) could
not be found, however, we checked that at least 40 approximants remain in
all cases.  For such sets of fewer than 80 approximants we increase our
uncertainty estimate by the ratio of the maximal number of Pad\'e approximants
(80) over the actual number in the set.

For space-like momenta our exclusion region is defined by
\begin{equation} 
  \text{Re}(z(\omega_0)) \leq 2 
  \quad\&\quad
  |\omega_0|\leq1.2\,,
  \label{eq:pole_crit_space}
\end{equation}
and negative values of $a_R$ in the range $[-10,-0.1]$ are chosen.


\section{\label{sec::res}Results}

\begin{figure}[t]
  \centering
  \includegraphics[width = 0.6\textwidth]{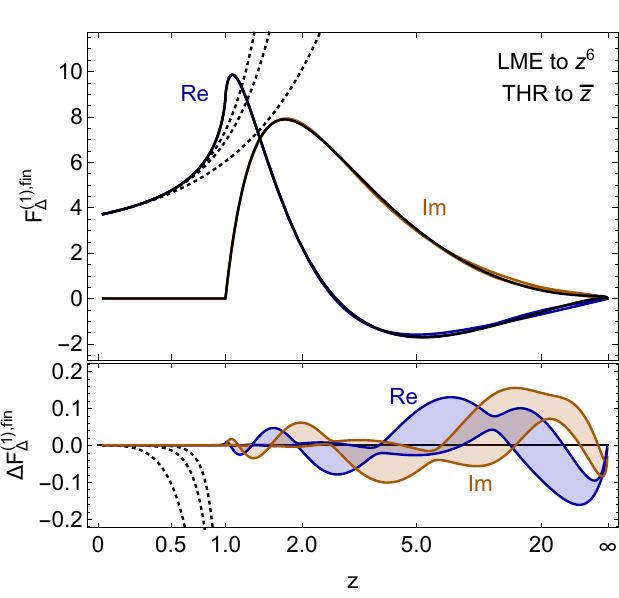}
  \caption{The upper panel shows our approximations for the real and imaginary 
    parts of the two-loop triangle form factor in blue and orange, respectively. The bands 
    give the standard deviation of the Pad\'e approximants which we consider. The exact 
    results are shown in black. The dashed lines correspond to the real part
    of the LME approximation up to order $z^2$, $z^4$ and $z^6$. 
    The lower panel shows the difference 
    between the exact result and the approximations.
    \label{fig:Ftri2Loop}}
\end{figure}

\begin{figure}[t]
  \centering
  \includegraphics[width=\textwidth]{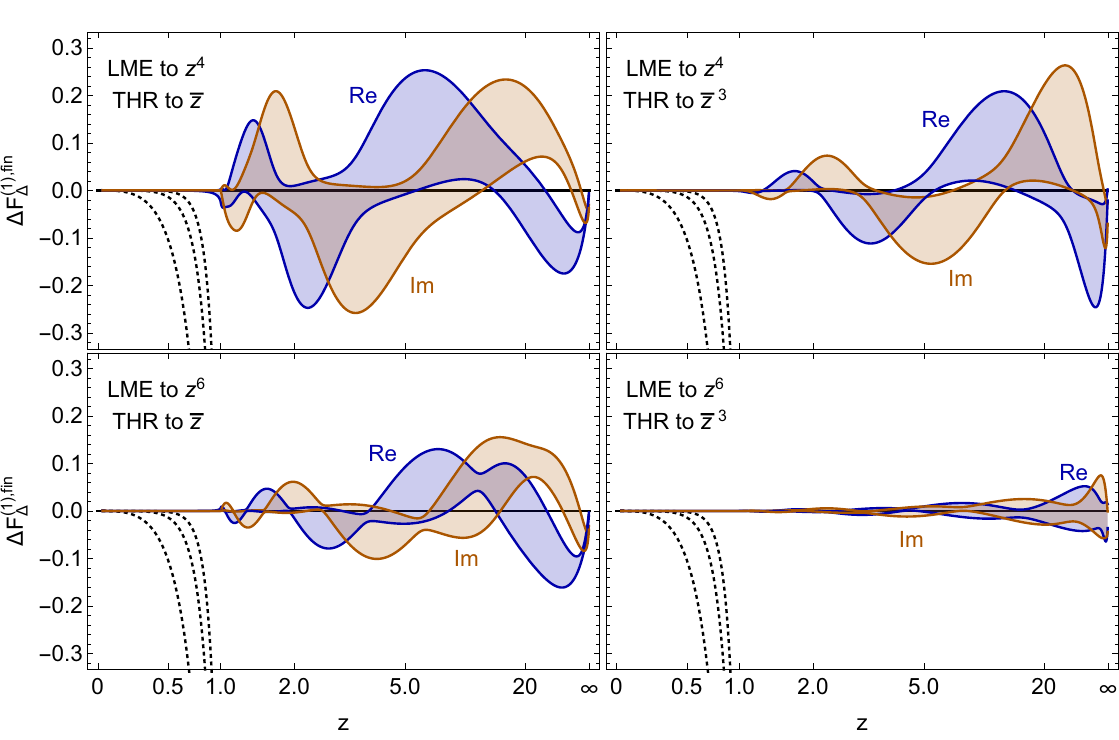}
  \caption{\label{fig:Ftri2LoopDiffs}Differences between our
      approximations and the exact result. The input used is shown in the
    legend of each panel.  The dashed lines correspond to the real part
    of the LME approximation up to order $z^2$, $z^4$ and $z^6$.  The lower
    left panel corresponds to the lower panel of Figure~\ref{fig:Ftri2Loop}.}
\end{figure}

Before discussing the three-loop results we apply the method described
in the previous section to the two-loop form factor, for which we can compare
to the exact
expressions~\cite{Spira:1995rr,Harlander:2005rq,Anastasiou:2006hc,Aglietti:2006tp}.

We show in Figure~\ref{fig:Ftri2Loop} that the exact result for the two-loop
form factor can be reproduced very well with the same amount of information
that is available at three loops. The shaded region is spanned by the standard
deviation w.r.t. to the mean value of a set of 20 approximants for each
considered set $\{n,m\}$. These approximants are available in the ancillary
file~\cite{progdata}.  Figure~\ref{fig:Ftri2LoopDiffs}, where the difference
between the exact result and the approximations is shown, demonstrates that
the approximation can be systematically improved by including more expansion
coefficients.  We compare the results based on the input used in
Figure~\ref{fig:Ftri2Loop} (lower left panel) to results where fewer expansion
coefficients for large top quark masses are used (upper left
panel). Furthermore, we also show results where additional information from
threshold is incorporated in the construction of the Pad\'e approximations
(panels on the right).

\begin{figure}[t]
  \centering
 \includegraphics[width = 0.6\textwidth]{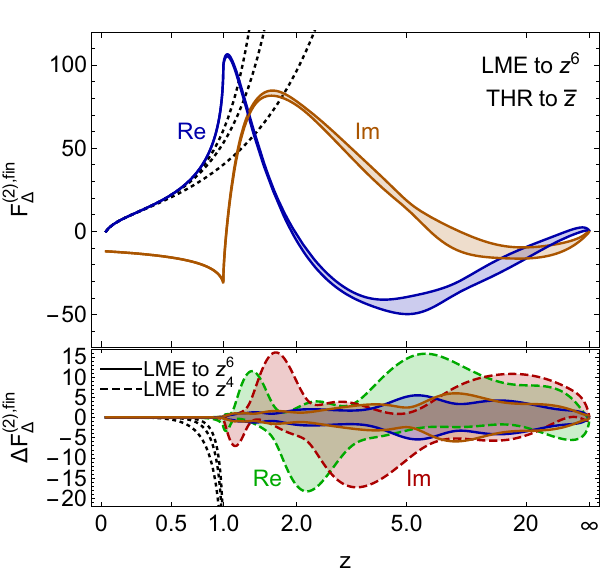}
 \caption{Our approximations for the real and imaginary parts of the
   three-loop triangle form factor are shown in blue and orange,
   respectively. The bands give the standard deviation of the considered set
   of Pad\'e approximants. The dashed lines correspond to the real part
     up to order $z^2$, $z^4$ and $z^6$.
   The lower panel shows the differences to the central values; see text
   for details.
   \label{fig:Ftri3Loop}}
\end{figure}

Our approximation of the three-loop form factor is shown in
Figure~\ref{fig:Ftri3Loop} and represents the main result of this paper. At
three loops the LME and threshold coefficients develop terms linear in
$L_s = \log(-4z-i0)$. We construct separate Pad\'e approximants for the
coefficient such that we obtain an approximation of the form
\begin{equation}
  F_\triangle^{(2),\rm fin}(z(\omega)) \simeq
    \frac{[n/m]_0(\omega) 
      + F_\triangle^{(2),\rm sub}}{1+a_{R,0} z(\omega)} 
    + \frac{[k/l]_1(\omega)L_s}{1+a_{R,1} z(\omega)}
    \,,
\label{eq:Ftri3l_Pade}
\end{equation}
where the subscripts indicate the power of $L_s$.  Note that the Pad\'e
approximants of the $L_s$-independent and linear-$L_s$ term are averaged
independently using separate values of $a_R$.  The threshold subtraction
(cf. Eq.~(\ref{eq::Fthrsub})) is only needed for the first term in
Eq.~(\ref{eq:Ftri3l_Pade}).  The lower panel in Figure~\ref{fig:Ftri3Loop}
shows the differences from the central values (obtained using seven expansion
terms for small $z$) both with seven and five input terms from the large top
quark mass expansion as solid and dashed boundaries of the uncertainty bands,
respectively. One observes over the whole range in $z$ (except for a small
region for $z\approx 10$) that the solid bands lie within the dashed band.
Below threshold ($z=1$) our method results in tiny uncertainties for
both the real and imaginary parts of the form factor. For $1\le z\le2$ the form
factor is numerically large and we thus observe small relative
uncertainties. Although the absolute uncertainty becomes larger for higher
values of $z$ we can provide a good approximation with an uncertainty which is
sufficiently small for phenomenological applications.

\begin{figure}
  \centering
  \includegraphics[width = 0.49\textwidth]{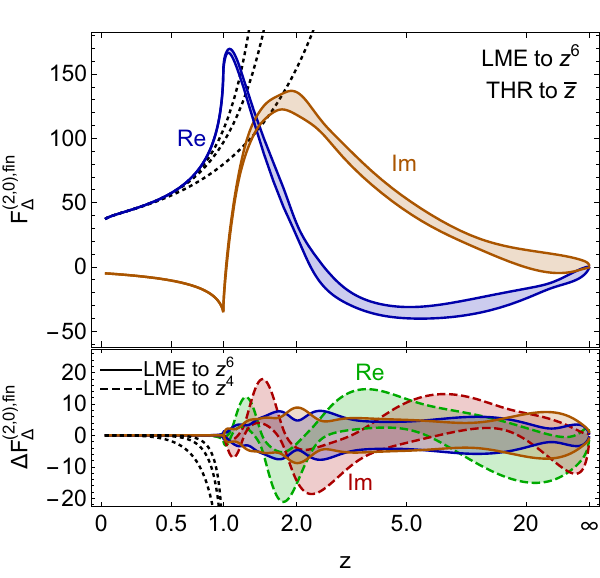}
  \hfill
  \includegraphics[width = 0.49\textwidth]{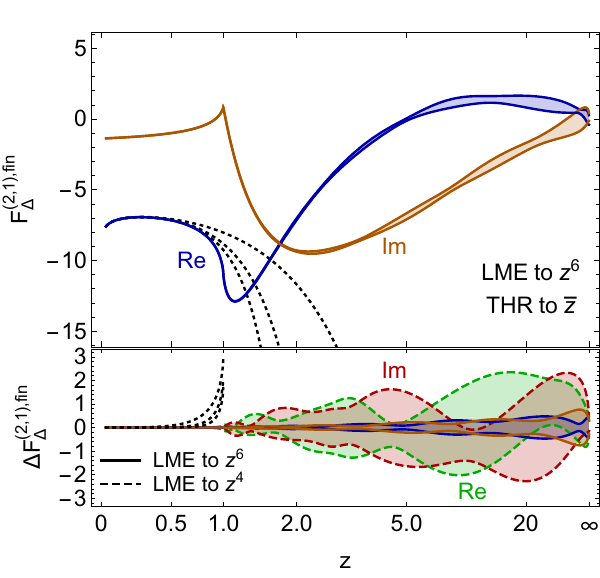} \hfill
  \caption{Our approximations for the three-loop form factor
    separated according to the light-fermion contributions.
    \label{fig:Ftri3LoopNf}}
\end{figure}

In order to facilitate the comparison with a future exact calculation we 
split our three-loop result according to the number of light fermions
and write
\begin{equation}
  F_\triangle^{(2),\rm fin}(z) = 
  F_\triangle^{(2,0),\rm fin}(z) 
  + n_l F_\triangle^{(2,1),\rm fin}(z) 
  + n_l^2F_\triangle^{(2,2),\rm fin}(z)\,, 
  \label{eq::split}
\end{equation}
where $n_l=5$ is the number of light flavors. Note that
$F_\triangle^{(2,0),\rm fin}(z)$ contains contributions with closed massive
loops, which are numerically less important than the $n_l$ terms. There are no
three-loop vertex diagrams which contain two closed fermion loops;
$F_\triangle^{(2,2),\rm fin}(z)$ is completely determined by the infra-red
subtraction terms. In fact, it is proportional to $F_\triangle^{(0)}$ and
we will not discuss it further.

The results for $F_\triangle^{(2,0),\rm fin}(z)$ and
$F_\triangle^{(2,1),\rm fin}(z)$ are shown in Figure~\ref{fig:Ftri3LoopNf},
adopting the notation from Figure~\ref{fig:Ftri3Loop}. Both coefficients show
a convergence which is very similar to $F_\triangle^{(2),\rm fin}$.  Summing
up the coefficients leads to good agreement with the
result~\eqref{eq:Ftri3l_Pade} but with a larger uncertainty which is
why~\eqref{eq:Ftri3l_Pade} should be used for numerical applications.  Note
that the heavy-quark contribution to $F_\triangle^{(2),\rm fin}(z)$ is
the only result where the exclusion criterion~(\ref{eq:pole_crit_2}) has been
used whereas for all other results~(\ref{eq:pole_crit}) is applied.

\begin{figure}[t]
 \centering
 \includegraphics[width = 0.6\textwidth]{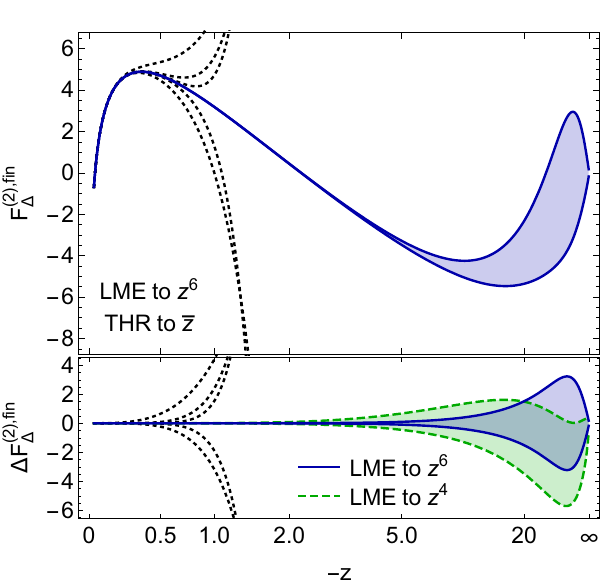}
 \caption{\label{fig:Ftri3Loop_spacelike}
   The three-loop result for the form factor $F_\triangle^{(2),\rm fin}$
   for space-like momenta. 
   The dashed lines correspond to the real part of the LME approximation up           
   to order $z^2$, $z^3$, $z^4$, $z^5$ and $z^6$.
   The same notation as in
   Figure~\ref{fig:Ftri3LoopNf} is adopted.}
\end{figure}

Finally, we present in Figure~\ref{fig:Ftri3Loop_spacelike} results for the
three-loop form factor for $z<0$. One observes small uncertainties for $|z|<5$
which become larger when $z$ becomes more negative.  For $|z|>20$ the
Pad\'e approximation procedure does not lead to accurate results, we note
that incorporating additional expansion terms does not significantly affect
the size of the uncertainty.  Note that the large top quark mass expansion
shows an alternating behaviour.

Together with this paper we provide representative Pad\'e approximants for all
plots shown in this section in an ancillary file~\cite{progdata}.


\section{\label{sec::con}Conclusion}

We compute three-loop corrections to the Higgs-gluon form factor including
finite top quark mass effects. Our approach is based on the combination of
analytic results from two kinematic regions: the expansion for large
top quark mass and the top quark threshold.  In addition, we incorporate
the information that the form factors vanish at high energies by a rescaling
(cf. Eq.~(\ref{eq:Pade_rescaling})). For the rescaled form factors, we apply
a conformal mapping and a subsequent Pad\'e approximation. We first 
apply our method at two loops and show that we can reproduce the known
results. The two-loop expression is also used to demonstrate that our estimate
for the uncertainty works reliably.  Our main result is shown in
Figure~\ref{fig:Ftri3Loop} where we plot the three-loop form factor in the
time-like momentum region.  This plot can be reproduced using the
approximation functions which are provided in the ancillary
file~\cite{progdata}.  We have shown that our results can be systematically
improved by incorporating more expansion terms into the analysis.



\section*{Acknowledgements}

RG is supported by the ``Berliner Chancengleichheitsprogramm''.
This research was supported by the Deutsche Forschungsgemeinschaft (DFG,
German Research Foundation) under grant 396021762 --- TRR 257
``Particle Physics Phenomenology after the Higgs Discovery''
and has received funding from the European Union's Horizon
2020 research and innovation programme under the Marie Sk{\l}odowska-Curie
grant agreement No. 764850, SAGEX.
We thank Robert Harlander, Mario Prausa and Johann Usovitsch for pointing
out a missing factor $1/2$ in Eq.~(\ref{eq::amp}) in the first version of the
manuscript.




\end{document}